# The effective fine structure constant of freestanding graphene measured in graphite

James P. Reed[1], Bruno Uchoa[1], Young Il Joe[1], Yu Gan[1], Diego Casa[2], Eduardo Fradkin[1], Peter Abbamonte[1]*

[1]Department of Physics and Frederick Seitz Materials Research Laboratory, University of Illinois, Urbana, IL, 61801, USA

[2]Advanced Photon Source, Argonne National Laboratory, 9700 S. Cass Av., Argonne, IL, 60439, USA

**Electrons in graphene behave like Dirac fermions, permitting phenomena from high energy physics to be studied in a solid state setting. A key question is whether or not these Fermions are critically influenced by Coulomb correlations. We performed inelastic x-ray scattering experiments on crystals of graphite, and applied reconstruction algorithms to image the dynamical screening of charge in a freestanding, graphene sheet. We found that the polarizability of the Dirac fermions is amplified by excitonic effects, improving screening of interactions between quasiparticles. The strength of interactions is characterized by a scale-dependent, effective fine structure constant, $\alpha_g^*(\mathbf{k},\omega)$, whose value approaches $0.14 \pm 0.092 \sim 1/7$ at low energy and large distances. This value is substantially smaller than the nominal $\alpha_g = 2.2$, suggesting that, on the whole, graphene is more weakly interacting than previously believed.**

*Email: abbamonte@mrl.uiuc.edu

Graphene is a single layer of carbon atoms with an unusual electronic structure that mimics the massless Dirac equation, allowing phenomena familiar from high energy physics to be investigated in a solid state setting (*1*). Because of its low density of states near the Fermi level, it is possible to tune the effective carrier density of graphene with a gate voltage. This makes graphene the potential foundation for a new generation of low-cost, flexible electronics (*1,2*).

It is widely believed that graphene, if isolated from substrate effects, should be a strongly interacting electron system.(*1, 3, 4, 5, 6, 7, 8*) The strength of Coulomb interactions in graphene is governed by the ratio of its potential energy to its kinetic energy, $U/K = e^2/\hbar v_F = 2.2$, where $e$ is the charge of an electron, $\hbar$ is Planck's constant, and $v_F$ is the Fermi velocity of the Dirac particles. This ratio is independent of the carrier density and usually referred to as the "fine structure constant", $\alpha_g$. Unlike the analogous quantity $\alpha = 1/137$ in quantum electrodynamics (QED), $\alpha_g$ is greater than unity; thus there is no small expansion parameter for electromagnetic interactions, which have been predicted to lead to novel ground states such as an excitonic insulator (*3*), or a perfect fluid that might exhibit electronic turbulence (*4*).

Surprisingly, so far there is little direct evidence for strong interactions in graphene. The hallmark of interactions is a logarithmically divergent renormalization of the Fermi velocity, $v_F$ (*5, 8*). However, this effect has not been observed in either angle-resolved photoemission (ARPES) experiments (*9, 10*) or in scanning single electron transistor (SET) measurements of the electronic compressibility (*11*). A recent optical infrared measurement observed a departure from the non-interacting spectrum (*12*), however the effect is not logarithmic and does not agree with ARPES or SET measurements. Interaction effects have been observed in high magnetic fields, but in this case the kinetic energy is quenched by the formation of Landau levels (*13, 14, 15*). Some of these measurements were done on supported graphene, which can suppress interactions through substrate dielectric screening. However, recent measurements show that free-standing graphene in zero field also behaves like a simple semimetal (*16*).

The absence of a $v_F$ renormalization seems irreconcilable with a large value of the fine structure constant. However, the particles measured in experiments are not bare electrons, but dressed quasiparticles, which interact via the screened Coulomb interaction (*17*). Hence, a better measure of the strength of interactions is the dressed fine structure constant, $\alpha_g^*(\mathbf{k},\omega) = \alpha_g / \varepsilon(\mathbf{k},\omega) = \alpha_g[1+V(k)\chi(\mathbf{k},\omega)]$, where $V(k) = 2\pi e^2/k$ is the bare Coulomb interaction in two dimensions, $\varepsilon(\mathbf{k},\omega)$ is the dielectric function, and $\chi(\mathbf{k},\omega)$ is the charge response function of graphene. Unlike $\alpha_g$, $\alpha_g^*(\mathbf{k},\omega)$ describes the retarded interaction among the dressed quasiparticles, and accounts for the influence of screening generated dynamically within the Dirac system (*18*). $\alpha_g^*(\mathbf{k},\omega)$ is not a "background" dielectric constant, but a parameter that accounts for the dynamically generated screening by the valence electrons. Diagrammatic calculations may be structured in powers of $\alpha_g^*(\mathbf{k},\omega)$, so this function can be considered a valid expansion parameter (*19*).

To determine $\alpha_g^*(\mathbf{k},\omega)$ one must determine the response function, $\chi(\mathbf{k},\omega)$, which is a general representation of the charge dynamics of the system. In real space, $\chi(\mathbf{r}_1-\mathbf{r}_2,t)$ represents the amplitude that a disturbance in the electron density at $\mathbf{r}_1$ will propagate to $\mathbf{r}_2$ after an elapsed time, $t$. $\chi(\mathbf{k},\omega)$ also describes, in linear response theory, how the system responds to charged perturbations via

$$n_{ind}(\mathbf{k},\omega) = V(\mathbf{k})\,\chi(\mathbf{k},\omega)\,n_{ext}(\mathbf{k},\omega), \qquad (1)$$

where $n_{ext}(\mathbf{k},\omega)$ is an arbitrary source and $n_{ind}(\mathbf{k},\omega)$ is the charge induced in the medium (*20*).

To determine $\chi(\mathbf{k},\omega)$ for graphene, we performed inelastic x-ray scattering (IXS) experiments on single crystals of graphite, which consists of layers of graphene loosely bound by electrostatic, Van der Waals forces. IXS measures the dynamic structure factor, $S(\mathbf{k},\omega)$, which is related to the imaginary part of the charge response through the quantum mechanical version of the fluctuation-dissipation theorem (*20, 21*),

$$S(\mathbf{k},\omega) = -\frac{1}{\pi}\frac{1}{1-e^{-\hbar\omega/kT}}\,\mathrm{Im}\!\left[\chi_{3D}(\mathbf{k},-\mathbf{k},\omega)\right], \qquad (2)$$

where $T$ is the temperature and $\chi_{3D}$ is the charge response of graphite. Eq. 2 provides only the imaginary part of $\chi_{3D}$; if a sufficiently broad energy spectrum is sampled in the experiment, the real part may be determined by Kramers-Kronig (KK) transform (*21,19*). We note that, because graphite is layered, its electron density is nonuniform, so $\chi_{3D} = \chi_{3D}(\mathbf{k},\mathbf{G}-\mathbf{k},\omega)$ is a function of two momenta, $\mathbf{G}$ being a reciprocal lattice vector. IXS experiments provide the $\mathbf{G}=0$ component (*21,19*).

The IXS spectra, shown in Fig. 1A,B (*19*), are dominated by two collective excitations, at approximately 6 eV and 35 eV, which were observed in previous studies (*22, 23*) and are normally referred to as the π and σ–π plasmons, respectively. These excitations are, however, not plasmons in the usual sense; they do not arise from free carrier screening, but from Van Hove singularities in the band structure. The former resides at the edge of the π bands near the M point in the Brillouin zone, and the latter in the σ-bonded bands along the A-L line (*23*). Beneath the plasmons, a continuum of single-particle excitations is visible. The spectra were measured, for $k_z$=0, throughout an entire, symmetry-independent sector of the Brillouin zone (Fig. 1C).

While these experiments were done on graphite crystals, the results are directly relevant to graphene. Graphite is a quasi-two-dimensional system in which the inter-layer hopping, $t_\perp = 0.4$ eV, is much smaller than that in-plane, $t_\parallel = 3$ eV. At energy scales greater than $t_\perp$, the π band of graphite is essentially the same as in graphene, and the system can be thought of as a stack of graphene sheets coupled only by direct, long-ranged Coulomb interactions.

More quantitatively, graphite and graphene have approximately the same polarizability, $\Pi(\mathbf{k},\omega)$. Physically, $\Pi$ can be thought of as the response of the system, excluding direct Coulomb interactions, which couple the layers, and is related to the response function by

$$\chi(\mathbf{k},\omega) = \frac{\Pi(\mathbf{k},\omega)}{1-V(k)\Pi(\mathbf{k},\omega)}. \qquad (3)$$

Making the assumption that $\Pi$ is the same for both graphite and graphene, and that the graphene sheets are very thin, we acquire an expression (*19*) for the response of graphene in terms of that for graphite,

$$\chi(\mathbf{k},\omega) = \frac{\chi_{3D}(\mathbf{k},-\mathbf{k},\omega)\cdot d}{1-V(k)\left[1-F(\mathbf{k})\right]\chi_{3D}(\mathbf{k},-\mathbf{k},\omega)\cdot d} \qquad (4)$$

where $d=3.35$ Å is the distance between the layers and

$$F(\mathbf{k}) = \frac{\sinh(qd)}{\cosh(qd)-\cos(k_z d)} \qquad (5)$$

is a form factor that describes the layered structure of graphite, $q$ and $k_z$ being the magnitude of the in- and out-of-plane components of $\mathbf{k}$, respectively (*19*). Eq. 4 can be thought of as a means of turning off the coulomb interaction between the layers, revealing the response for half-filled, freestanding graphene.

To test Eq. 4 we applied it to an electron energy loss (EELS) study by Eberlein (*24*) of the dielectric loss function $-\text{Im}[1/\varepsilon(\mathbf{k},\omega)]$, which is proportional to $\text{Im}[\chi]$, for both graphite and freestanding graphene. They found that the π and σ–π plasmons are also present in graphene, but are shifted to lower energy compared to graphite because of the reduced dimensionality (*23, 24*). We scaled their graphite spectra with the f sum rule, KK transformed, applied our Eq. 4, and compared the results to their spectra for graphene (Fig. 1D). The curves match nearly exactly, reproducing the red shifts and changes in spectral weight of the two plasmons. We conclude that Eq. 4 provides a valid response function for freestanding graphene at energy scales greater than $t_\perp$ (*19*).

At the lowest momenta measured, the $\chi(\mathbf{k},\omega)$ derived from Eq. 4 shows signatures of the Dirac fermions. In Fig. 2A we plot $-\text{Im}[\chi(\mathbf{k},\omega)]$ as a function of $\omega$ for two momenta at which the quantity $\hbar v_F k$ is less than the energy of the Van Hove singularity in the π band. Also shown is the spectrum for idealized, noninteracting Dirac fermions. A continuum is visible below 5 eV whose magnitude and dispersion with $\mathbf{k}$ are close to that expected for Dirac particles. This suggests that the low frequency, long wavelength response of graphene is strongly influenced by the Dirac fermions. The experimental and idealized spectra differ, however, in two respects. First, the curves deviate near the energy of the π plasmon, since in the real material the width of the π band is finite. Second, the experimental onset energy is lower and smoother than expected.

The origin of these discrepancies are clarified by examining the polarization function, $-\text{Im}[\Pi(\mathbf{k},\omega)]$, shown in Fig. 2B, which exhibits peaks at the energy of single-particle transitions rather than of the collective modes. Again, the curves for idealized Dirac fermions are shown. The peak in the data is broader and shifted to lower energy, by approximately 0.6 eV, compared the idealized spectrum. Further, the spectral weight in the Van Hove singularity is suppressed from what is expected (*23*), being reduced to a shoulder on the Dirac spectrum. These effects may be understood as arising from a combination of band curvature (i.e. deviation from the Dirac spectrum at high energy), as well as excitonic effects, which are known to transfer spectral weight to lower energy in particle-hole spectra. We note that the 0.6 eV shift is similar in magnitude to that predicted by recent *ab initio* calculations employing the Bethe-Saltpeter equation (*25*, *26*).

To observe how excitations in graphene propagate in real time, in Fig. 3A-F we plot the time-dependent charge density, $n_{ind}(\mathbf{r},t)$, arising from a point source in graphene, determined by using $n_{ext}(\mathbf{k},\omega) = -e$ in Eq. 1. As expected from causality, $n_{ind} = 0$ for $t < 0$. The earliest response ($t < 200$ *as*) is dominated by the fast, broadband, σ–π plasmon, which is isotropic and ~5 Å in size because of the localized character of the *sp*² bonds. The narrower π plasmon emerges later ($t > 200$ as), extends over ~10 Å, and exhibits six-fold rotational symmetry. One can think of the sequence of events as a localized burst of density that back-scatters off the atomic lattice.

The effective fine structure constant, $\alpha_g^*(\mathbf{k},\omega)$, is shown in Fig. 4. $\alpha_g^*(\mathbf{k},\omega)$ is a valid coupling constant at low energy where the bands are Dirac-like; at high energy it may be thought of simply as the inverse dielectric function, scaled by $\alpha_g$. Fig. 4 shows that $\alpha_g^*$ is complex, and is not a constant but is a strong function of frequency and momentum. This indicates that the strength of interactions in graphene depends on the scale on which the system is probed. The magnitude of $\alpha_g^*$ ranges from greater than 2 at high energy to significantly less than unity at energies lying below the $\omega = \hbar v_F k$ line (Fig. 4).

An important limit is that of zero frequency and small momentum. At the lowest momenta measured, we found that $\Pi(\mathbf{k},0) \sim k$ as $k \to 0$ (*19*). Extrapolating linearly to zero, we find that $\alpha_g^*(0^+,0) \equiv \lim_{\mathbf{k}\to 0} \alpha_g^*(\mathbf{k},0) = 0.14 \pm 0.092 \approx 1/7$, which may be thought of as a static dielectric constant of $\varepsilon = [1-Q(\infty)/e]^{-1} = 15.4^{+39.56}_{-6.45}$ (*19*). This large value, which is an outcome of the excitonic shifts shown in Fig. 2B, is 3.5 times larger than past estimates based on the Random Phase Approximation (RPA) (*27*) or *GW* methods (*26*) in which excitonic effects were neglected. The small value of $\alpha_g^*$ in this limit indicates that graphene can screen very effectively over finite distances, and should act like a weakly interacting system for phenomena that take place at low energy and modest wave vector.

For illustration, we simulate the response of the system to a charged impurity by solving Eq. 1 with $n_{ext}(\mathbf{k},0) = -e$. The induced charge $n_{ind}(\mathbf{r})$ (Fig. 3G) is a cloud approximately 10 Å in size, comprising an isotropic inner layer surrounded by a six-fold symmetric outer layer, reminiscent of the excitations in Fig. 3A-F. Fig. 3H shows $Q(R)$, which is the total charge contained in $n_{ind}(\mathbf{r})$ within a shell of radius $R$. $Q(R)$ oscillates and plateaus at large $R$ to the value $Q(\infty) = (0.924 \pm 0.046)e$. Hence, at length scales larger than a few nm, a charged impurity in graphene is screened nearly completely. This explains the surprisingly small sensitivity of the mobility of graphene to a high-κ dielectric environment (*28, 29*), which was anticipated to reduce scattering from charged impurities. Because graphene already screens impurities very efficiently, immersing it in a high-κ environment does not significantly influence its properties. Screening effects may also explain the absence of a detectable velocity renormalization in recent ARPES and SET experiments. Overall, our results indicate that graphene is a more weakly interacting system than previously assumed, and that the mobility of graphene may currently be limited by some phenomenon other than scattering from charged impurities.

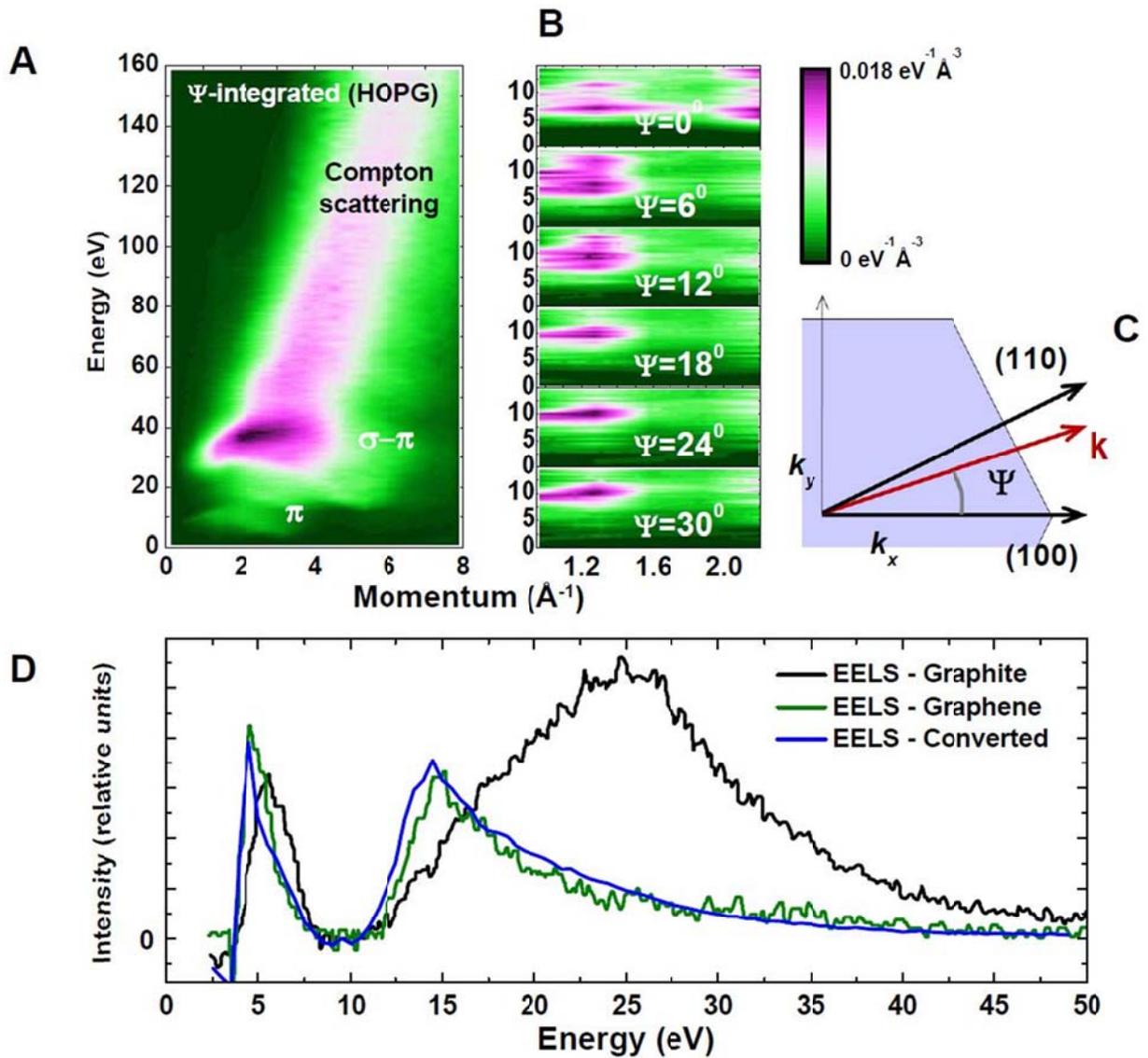

**Figure 1** Inelastic x-ray scattering experiments from graphite and extraction of the response function for graphene. (*a*) Scattered intensity as a function of energy and momentum for highly oriented pyrolitic graphite (HOPG), which gives the $\Psi$-integrated response. (*b*) Angle-resolved spectra from single crystal graphite for the domain over which anisotropy was observed. (*c*) Brillouin zone of graphene with various vectors defined. (*d*) Test of Eq. 4 on the electron energy loss experiments of Eberlein (*24*) (fit value $k = 0.33$ Å$^{-1}$), showing that an accurate response for graphene can be obtained from IXS experiments on graphite.

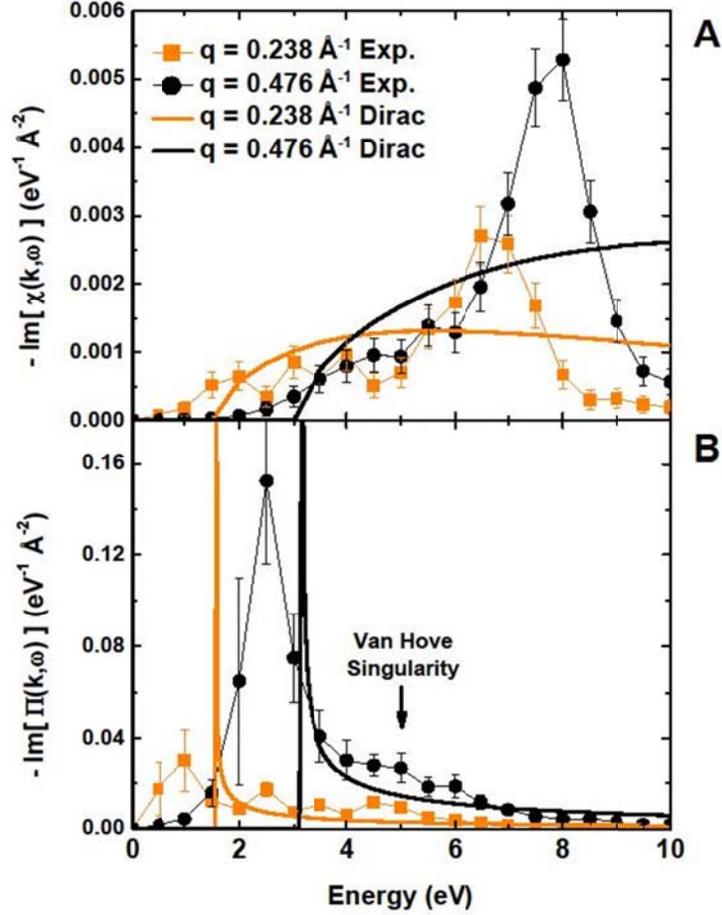

**Figure 2** Response functions for graphene in the low-momentum region, showing the Dirac particles. (*a*) $-\text{Im}[\chi(\mathbf{k},\omega)]$. The Dirac spectrum is visible as a continuum below the π plasmon, which arises from the Van Hove singularity at the top of the π band. Also shown is the expected response for idealized, noninteracting Dirac fermions, using Eq. 3 and $\Pi(\mathbf{k},\omega) = -q^2 / \left(4\sqrt{(\hbar v_F q)^2 - \omega^2}\right)$, with $v_F = 10^6$ m/s. The curves are plotted on an absolute scale and the magnitudes may be directly compared. (*b*) Experimental and idealized $-\text{Im}[\Pi(\mathbf{k},\omega)]$, showing the locations of single-particle excitations. The peak in the measured response is 0.6 eV lower in energy than the ideal response, indicating excitonic effects, which transfer spectral weight from the Van Hove singularity to the Dirac spectrum.

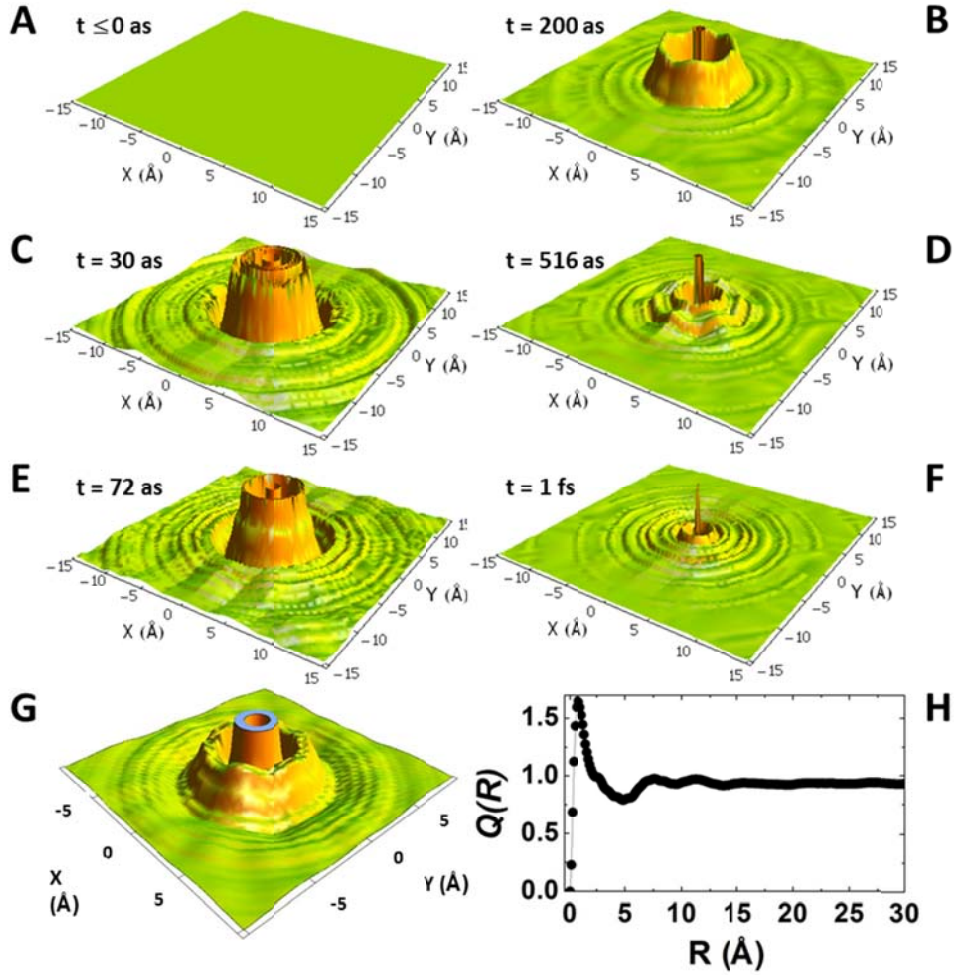

**Figure 3** Dynamics of the valence electrons in graphene. (*a–f*) Dynamical electron density, $n_{ind}(x,y,t)$, induced by a point source appearing at the origin and then instantaneously disappearing. We have taken $e$=1. Densities are plotted over a field of view of 30 Å × 30 Å at selected times. The vertical scale has been truncated at $\pm 5.0$ Å$^{-2}$ $as^{-1}$ to magnify the finer features. The resolutions for these images were $\Delta \mathbf{r}$ = 0.20 Å and $\Delta t$ = 10.3 *as*. The time resolution is fast in comparison to currently available laser pulses. (*g*) Time-independent electron density, $n_{ind}(x,y)$, induced by a static point charge. The vertical scale has been truncated at $\pm 0.04$ Å$^{-2}$ to magnify the finer features. (*h*) Integrated density $Q(R)$, giving an asymptotic value of $Q(\infty) = (0.924 \pm 0.046)e$, showing that a static point charge in graphene is screened nearly completely.

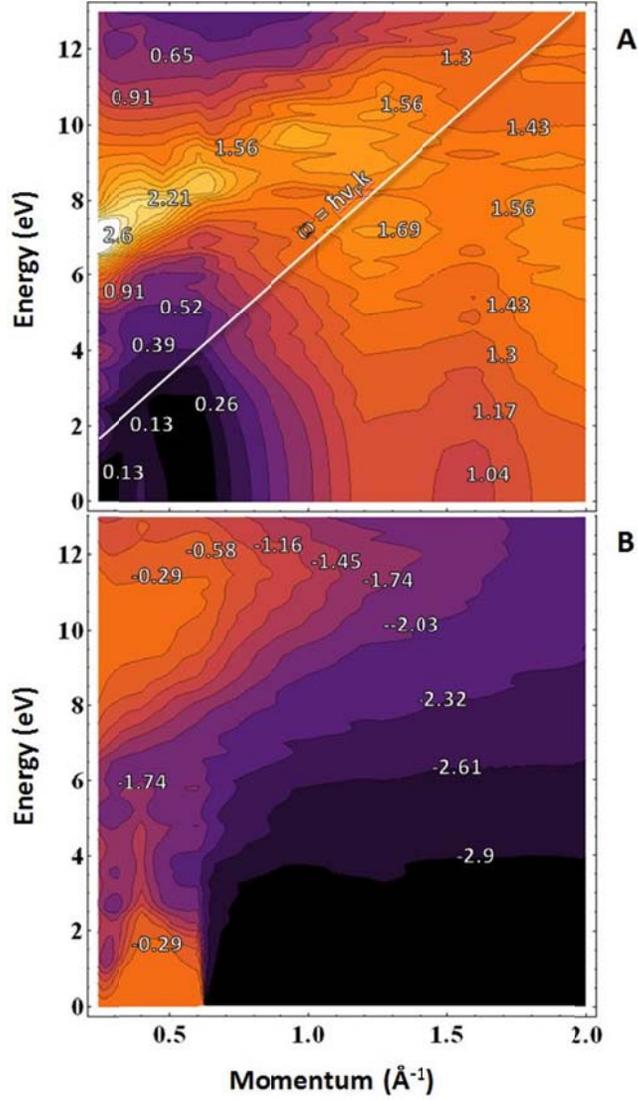

**Figure 4** The effective, screened fine structure constant, $\alpha_g^*(\mathbf{k}, \omega)$, as defined in the text. (*a*) The magnitude of $\alpha_g^*(\mathbf{k}, \omega)$, plotted against momentum and energy. The Dirac dispersion $\omega = \hbar v_F k$ is indicated by the white line. In the low momentum region, $\alpha_g^*(\mathbf{k}, \omega)$ is larger above this line than below. (*b*) The phase of $\alpha_g^*(\mathbf{k}, \omega)$, in radians.

**Supporting Online Material**
www.sciencemag.org
Materials and Methods
Figs. S1 – S6
Movie M1, M2

Supporting online material for

# The effective fine structure constant of freestanding graphene

James P. Reed[1], Bruno Uchoa[1], Young Il Joe[1], Yu Gan[1], Diego Casa[2],
Eduardo Fradkin[1], Peter Abbamonte[1]*

*To whom all correspondence should be addressed.  E-mail: abbamonte@mrl.uiuc.edu

**This PDF file includes:**

Animation of the densities, M1, M2

Materials and Methods

Supporting text

Figs. S1-S6

References S1-S8

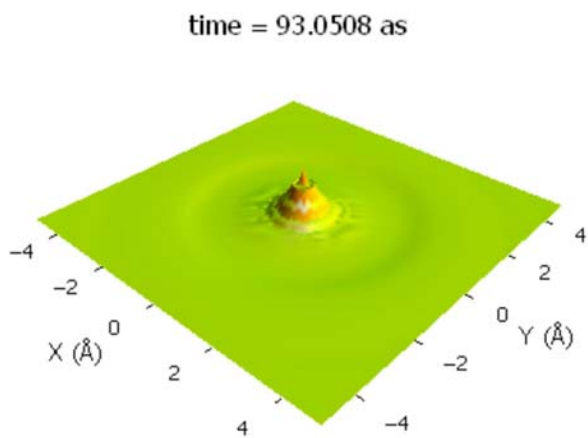 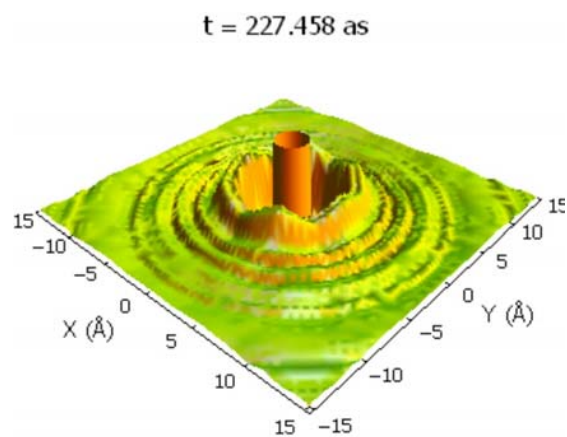

**M1**  Animation of $n(x,y,t)$ displayed in Fig. 2 at two different vertical scales.
http://users.mrl.uiuc.edu/abbamonte/density_graphene.avi

**M2**  Same animation of $n(x,y,t)$ as in M1 displayed on an expanded vertical scale to reveal finer features.
http://users.mrl.uiuc.edu/abbamonte/density_closeup.avi

## Materials and Methods

*IXS measurements and data processing*

Inelastic x-ray scattering (IXS) experiments were done at Sector 9-ID (XOR) at the Advanced Photon Source. Measurements were done in transmission geometry on commercially obtained single crystal graphite (SCG), as well as ZYA grade, highly oriented pyrolytic graphite (HOPG). A Ge(733) analyzer was used near backscattering with $E_i = 8980.5$ eV. Combined with the use of a Si(333) secondary monochromator, our energy resolution was $\Delta E = 0.3$ eV. A sample vessel with an in-vacuum, motorized beam stop allowed spectra to be obtained for scattering angles over the range $3° < 2\theta < 120°$, i.e. momentum transfers in the range $0.238 \text{Å}^{-1} < |\mathbf{k}| < 7.88 \text{Å}^{-1}$. An in-vacuum goniometer head allowed the azimuthal angle $\Psi$ to be adjusted over a 90° range. Spectra were taken over the transferred energy range $-3 < \omega < 200$ eV.

The elastic scattering was removed from individual spectra by subtracting a Voigt fit to the zero-loss line. The continuum features highlighted in Fig. 2A were visible in the raw data on the tail of the elastic scattering in two separate experiments, so were not an artifact of the subtraction. The spectra were corrected for changes in scattering volume, and placed on an absolute scale with the F sum rule (*S1*),

$$\int_0^\infty d\omega\, \omega\, \text{Im}\left[\chi_{3D}(k,-k,\omega)\right] = \frac{\pi n k^2}{2m} \tag{S1}$$

This integral was satisfied at many momenta without any scaling required among the different *k* values (Fig. S1). The error bars on various quantities, such as the effective charge and fine structure constant, were determined by taking a 5% overall variance in the magnitude of the integral in Eq. S1. At momenta higher than 7.87 Å$^{-1}$ the spectra consist of simple Compton scattering. This allowed the data to be extrapolated to $\omega \to \infty$ and $k \to \infty$ by scaling the Compton profile. The spectra were found to be independent of the azimuthal angle, with the exception of the π plasmon, which was quite anisotropic over the range 0.9 Å$^{-1}$ < *k* < 2.2 Å$^{-1}$ (Fig. 1B). To get a full data set we collected data from the SCG sample over this momentum range and combined it with data from HOPG to construct the quantity $\text{Im}[\chi(k_x, k_y, 0, \omega)]$ over a symmetry-independent, 30° sector of the Brillouin zone (Fig. 1C). This quantity was then Kramers-Kronig transformed to acquire the real part, $\text{Re}[\chi(k_x, k_y, 0, \omega)]$, and repeated to construct a complete, two-dimensional parameterization of the complex $\chi(k_x, k_y, 0, \omega)$.

*Construction of charge densities*

Data analysis was performed with Mathematica 7.0. Because the experiments were done on graphite crystals, we begin by determining the charge densities that arise from either a static or dynamic point source in three-dimensional graphite. The central challenge is dealing with the issue, mentioned in the main manuscript, that the complete response of graphite, $\chi_{3D}(\mathbf{k}, \mathbf{G}-\mathbf{k}, \omega)$, is a function of two distinct momenta, the appropriate generalization of Eq. 1 being

$$n_{ind}(\mathbf{k}, \omega) = \sum_{\mathbf{G}} \chi_{3D}(\mathbf{k}, \mathbf{G}-\mathbf{k}, \omega) V_{3D}(\mathbf{G}-\mathbf{k}) n_{ext}(\mathbf{k}, \omega), \tag{S2}$$

where $V_{3D}(k) = 4\pi e^2 / k^2$ is the Coulomb propagator in three dimensions. IXS probes only the **G**=0 component, $\chi_{3D}(\mathbf{k}, -\mathbf{k}, \omega)$ (*S2*). In a homogeneous system only the **G**=0 term is nonzero, so this experimental restriction poses no limitation. However graphite is highly layered, so it is necessary to make some assumption about how charge is arranged in the system, i.e. how $\chi_{3D}(\mathbf{k}, \mathbf{G}-\mathbf{k}, \omega)$ depends on **G**.

Rather than appealing to a microscopic model, we will consider two limiting cases. The first is the homogeneous limit, in which the charge density is assumed to be uniform in space, and only the **G**=0 term in Eq. S2 is used. The second is the layered limit, in which the graphene sheets are assumed to be arbitrarily thin, in which case $\chi_{3D}(\mathbf{k}, \mathbf{G}-\mathbf{k}, \omega)$ is independent of the component $G_z$ and nonzero only when $G_x = G_y = 0$. In this limit, $\chi$ may be factored out of the sum in Eq. S2, which can be done analytically, as shown in Eq. S3. The layered case is a decidedly more realistic representation of real graphite than the homogeneous case, but we will consider both limits as a means of placing bounds on the properties of the real system.

First we consider the homogeneous limit. To acquire the dynamical response we first constructed the quantity $\text{Im}[n_{ind}(k_x, k_y, \omega)] = -eV_{3D}(k) \text{Im}[\chi_{3D}(k_x, k_y, 0, \omega)]$, which is the imaginary part of the **G** = 0 term in Eq. S2, with $n_{ext}(\mathbf{k}, \omega) = -e$. The quantity $\text{Im}[\chi_{3D}(k_x, k_y, 0, \omega)]$ was found to have a quadratic dependence on the magnitude of **k** at small values of **k** (Fig. S1), so was extrapolated quadratically to **k** = 0. It was interpolated onto a continuous frequency interval, which is necessary for acquiring causal images (*20*). The quantity $n_{ind}(x, y, t)$ was then acquired by performing a frequency sine transform of $\text{Im}[n_{ind}(k_x, k_y, \omega)]$ and Fourier transforming the momentum variables. The advantage of using a sine transform is that explicit Kramers-Kronig transformation is not required (*S3*). The result is shown in Fig. S2. We note that these images were constructed from data with fixed $k_z = 0$, so represent the three-dimensional response, $n_{ind}(x, y, z, t)$, integrated along *z* (perpendicular to the graphene sheets). The resolutions of these images, defined in terms of the Nyquist theorem (*S4*), are $\Delta t = 10.3$ *as* and $\Delta r = 0.20$ Å. We note that our time resolution is shorter than the fastest laser pulse ever produced (*S5*). As required by causality, $n_{ind} = 0$ for $t < 0$. The earliest response ($t < 200$ *as*) is dominated by the fast, broadband, σ–π plasmon, which is isotropic and only ~5Å in size because of the localized character of the $sp^2$ bonds. The π plasmon, which has a narrower linewidth, emerges later ($t > 200$ as), extends over ~15 Å, and exhibits six-fold rotational periodicity. One can think of the sequence of events as a rotationally symmetric burst of density that back-scatters off the atomic lattice.

To determine how these processes dress a static charge, we solved Eq. S2 for a static source in graphite, $n_{ext}(\mathbf{k},0) = -e$. We performed a single-energy Kramers-Kronig transform to acquire $\chi_{3D}(k_x,k_y,0,0)$, which is purely real, from $\text{Im}[\chi_{3D}(k_x,k_y,0,\omega)]$, and constructed $n_{ind}(k_x,k_y) = -eV_{3D}(k)\chi_{3D}(k_x,k_y,0,0)$, which is the $\mathbf{G} = 0$ term in Eq. S2 for the static case, and Fourier transformed. The resulting $n_{ind}(\mathbf{r})$ (Fig. S3A), reminiscent of the excitations in Fig. S2, is ~15 Å in size and comprises an isotropic inner layer surrounded by a six-fold symmetric outer layer.

Next, we acquired the charge density in the limit that the layers are assumed to be arbitrarily thin. In this limit, the quantity $\chi_{3D}(\mathbf{k},\mathbf{G}-\mathbf{k},\omega)$ will be independent of $G_z$, where $\mathbf{G} = (G_x, G_y, G_z)$ is a reciprocal lattice vector. Further, if each of the layers is homogeneous in the $(x,y)$ plane, $\chi_{3D}(\mathbf{k},\mathbf{G}-\mathbf{k},\omega)$ will be nonzero only for $G_x = G_y = 0$. With these assumptions, using again $n_{ext}(\mathbf{k},0) = -e$, Eq. S2 reduces to

$$n_{ind}(\mathbf{k}) = -e\chi_{3D}(\mathbf{k},-\mathbf{k},0)\sum_n \frac{4\pi e^2}{q^2 + (k_z - 2\pi n/d)^2} \tag{S3}$$

Where $q = \sqrt{k_x^2 + k_y^2}$ is the magnitude of the in-plane component of the momentum, $k_z$ is the out-of-plane component, $G_z = 2\pi n/d$, and $n$ is an integer. The sum on the right hand side can be done exactly, giving $n_{ind}(\mathbf{k}) = -e\chi_{3D}(\mathbf{k},-\mathbf{k})V(\mathbf{k})F(\mathbf{k}) \cdot d$, where $d$ is the distance between the layers, $V(k) = 2\pi e^2/k$, and $F(\mathbf{k})$ is given in Eq. 5 of the main manuscript (see also Eq. S6 below). The resulting function, $n_{ind}(k_x,k_y)$, was Fourier transformed to obtain Fig. S3B.

## Supporting Text

***Expression of the graphene response, $\chi(\mathbf{k},\omega)$, in terms of that for graphite, $\chi_{3D}(\mathbf{k},-\mathbf{k},\omega)$.***

In this section we derive Eq. 4 of the main manuscript. For a periodic system in any number of dimensions, the general relationship between the susceptibility, $\chi$, and the polarization function, $\Pi$, is

$$\chi(\mathbf{k},\mathbf{G}-\mathbf{k},\omega) = \Pi(\mathbf{k},\mathbf{G}-\mathbf{k},\omega) + \sum_{\mathbf{G}'} \Pi(\mathbf{k},\mathbf{G}'-\mathbf{k},\omega) V(\mathbf{k}-\mathbf{G}') \chi(\mathbf{k}-\mathbf{G}',\mathbf{G}-\mathbf{k},\omega) \quad (S4)$$

where $\mathbf{G}$ is a reciprocal lattice vector and $V(\mathbf{k})$ is the Coulomb propagator in the appropriate number of dimensions (the diagrammatic expression of this relationship is shown in Fig. S4A). For the case of graphite, if we assume, as above, that the layers are arbitrarily thin, then $\chi_{3D}$ as well as $\Pi_{3D}$ are both independent of the z-component, $G_z$. In this approximation, both $\chi_{3D}$ and $\Pi_{3D}$ may be factored out of the sum in Eq. S4, which can be done analytically in the same manner as in Eq. S3. The result is

$$\chi_{3D}(\mathbf{k},-\mathbf{k},\omega) = \frac{\Pi_{3D}(\mathbf{k},-\mathbf{k},\omega)}{1-V(\mathbf{k})F(\mathbf{k})\Pi_{3D}(\mathbf{k},-\mathbf{k},\omega)\cdot d} \quad (S5)$$

where $V(k) = 2\pi e^2 / k$ is the Coulomb propagator in two dimensions and the form factor

$$F(\mathbf{k}) = \frac{\sinh(qd)}{\cosh(qd) - \cos(k_z d)} \quad (S6)$$

describes how charge interacts in three dimensions. The product $V(\mathbf{k})F(\mathbf{k})\cdot d$ is essentially the Coulomb propagator for a layered system (*S6*).

For two-dimensional graphene, which we assume to be homogeneous within the plane, the corresponding relationship is

$$\chi(\mathbf{k},\omega) = \frac{\Pi_{2D}(\mathbf{k},\omega)}{1-V(\mathbf{k})\Pi_{2D}(\mathbf{k},\omega)}. \quad (S7)$$

If we assume the graphene layers in graphite are coupled only by direct Coulomb interactions, the polarization functions for graphite and graphene are the same, i.e.,

$$\Pi_{3D}(\mathbf{k},\omega) = \frac{1}{d}\Pi_{2D}(\mathbf{k},\omega) \quad (S8)$$

where the factor $1/d$ accounts for the different units. Using this relationship, combining Eq. S5 and Eq. S7, we get

$$\chi(\mathbf{k},\omega) = \frac{\chi_{3D}(\mathbf{k},-\mathbf{k},\omega) \cdot d}{1 - V(k)\left[1 - F(\mathbf{k})\right]\chi_{3D}(\mathbf{k},-\mathbf{k},\omega) \cdot d}, \quad (S9)$$

which is Eq. 4 of the main manuscript.

At energy scales where the hopping between the layers can be neglected, Eq. S8 is justified by the fact that the internal Coulomb lines that dress a polarization bubble, Π, can only scatter electrons within a given layer (Fig. S4B). In this regime, the layers are coupled only by direct Coulomb lines connecting different polarization bubbles, as shown in Fig. S4A. Because the electronic structure of graphite and graphene are very similar, Π($\mathbf{k}$,ω) can be expected to be roughly the same for the two systems at energy scales greater than $t_\perp$. Use Eq. S9 is therefore justified even in the presence of strong correlations, and in particular is justified beyond the range of validity of most common approximation schemes, such as RPA.

In the main manuscript the conversion formula Eq. S9 was tested on an electron energy loss (EELS) experiment by Eberlein (*24*). In Fig. S4C we also test this expression on a density functional theory calculation from the same article. We KK transformed their calculated graphite spectra, scaled it with the f sum rule, applied our Eq. S9, and compared the results to their calculated spectra for graphene (Fig. S4C). Again, the curves match exactly in the high energy region, reproducing both the red shifts and changes in spectral weight of the two plasmons. The small peak at 2 eV is not reproduced, but this feature is not present in their experimental data so may be an artifact of the *ab initio* calculation.

We note that at large momenta, i.e. $q \gg 1/d \sim 0.3\,\text{Å}^{-1}$, the form factor $F(\mathbf{k}) \to 1$, graphite behaves as a two-dimensional system, and the conversion Eq. S9 has no effect on the response function. In applying Eq. S9 it was found, indeed, to have no effect on the response for momenta $q > 1.2\,\text{Å}^{-1}$.

*Asymptotics*

The graphene response $\chi(\mathbf{k},\omega)$ was found to have fundamentally different asymptotic properties from the three dimensional $\chi_{3D}(\mathbf{k},-\mathbf{k},\omega)$. This is evident in Fig S5, which shows the momentum dependence of the zero-frequency values $\chi(\mathbf{k},0)$ and $\chi_{3D}(\mathbf{k},-\mathbf{k},0)$ in the low momentum region. Notice that, while $\chi_{3D}(\mathbf{k},-\mathbf{k},0) \sim k^2$ at small values of $\mathbf{k}$, $\chi(\mathbf{k},0) \sim k$. The difference is an outcome of application of Eq. S9, and reflects the underlying property that the polarization function $\Pi_{3D}(\mathbf{k},\omega) \sim k$ at small $\mathbf{k}$ for both graphite and graphene, which is expected for a system of Dirac particles.

The linearity of $\chi(\mathbf{k},0)$ in the long wavelength, asymptotic limit has the significant implication that graphene can support dielectric screening, which is not normally possible in two dimensions. In any number of dimensions, the dielectric function is related to the polarization by

$$\varepsilon(\mathbf{k},\omega) = 1 - V(\mathbf{k})\Pi(\mathbf{k},\omega).\tag{S10}$$

For a system with an energy gap, $\Pi(\mathbf{k},0) \sim k^2$ at small $\mathbf{k}$. In two-dimensions, $V(\mathbf{k}) \sim 1/k$, which indicates that $\varepsilon(\mathbf{k},0) \to 1$ as $\mathbf{k} \to 0$, implying that dielectric screening is not possible at large distances in two dimensions.

Graphene, however, is a special case. Because $\Pi(\mathbf{k},0) \sim k$ at small $\mathbf{k}$, Eq. S10 converges to a finite dielectric constant at large distances, meaning that graphene – because of its gapless electronic structure – can support dielectric screening. For this reason, when computing charge densities, the final graphene response was extrapolated linearly in momentum to $\mathbf{k} = 0$.

*Comparison of various charge densities*

Having determined the response for graphene, it is useful to compare the screening of a point charge for the three cases we have considered, namely graphite in the homogeneous limit (Fig. S3A), graphite in the layered limit (Fig. S3B), and graphene (Fig. 3G in the main manuscript). The density clouds in each of these figures acts to screen the external charge, so a quantity of particular interest is the total amount of charge contained within each of these clouds.

In Fig. S6 we plot $Q(R)$, which is the total charge contained within $n_{ind}(\mathbf{r})$, integrated out to a cutoff radius, $R$, for all three cases. The curves for graphene and layered graphite are nearly identical, and differ from the homogeneous graphite response in that they exhibit a region from 0.3 Å to 2.5 Å in which the net induced charge exceeds the value of the external charge. This "overscreening" region is an outcome of the assumption of infinitely thin layers, and indicates that – at extremely short distances – the Coulomb interaction is actually attractive. Because this feature does not occur if one assumes a homogeneous charge density, our results do not unequivocally prove that it is real. However since the real system is highly layered it is reasonable to conjecture that overscreening may occur, at least to some degree.

The asymptotic value of the total charge, $Q(\infty)$, however, is roughly the same in the three cases. In the homogeneous limit, the integrated charge plateaus at large distance to the value $Q(\infty) = (0.924 \pm 0.046)e$. In the layered case $Q(\infty) = (0.935 \pm 0.046)e$. In the graphene case, $Q(\infty) = (0.924 \pm 0.046)e$. These three values give a long-wavelength dielectric constants $\varepsilon = [1 - Q(\infty)/e]^{-1} \sim 15$ in all three cases. In graphene, this suggests a zero frequency, renormalized fine structure constant of $\alpha_g^* = e^2/\varepsilon\hbar v_F = 0.142 \pm 0.092 \approx 1/7$.

We emphasize that in all cases – both for graphite and graphene – our analysis has made the assumption of linear response. In this approximation, we find that the screening is substantial, i.e., when integrated out to a modest cutoff the residual net charge of an electron is only a small fraction of *e*. Nonetheless, the value of the density at any given location is small compared to the background charge density, indicating *a posteriori* that non-linear corrections to the response are small.

***Relationship between $\alpha_g^*(\mathbf{k},\omega)$ and physical observables, such as the self-energy, $\Sigma^*(\mathbf{k},\omega)$***

In any many-body system, the one electron Green's function is defined as

$$G(k) = \frac{G^0(k)}{1-\Sigma^*(k)G^0(k)}, \tag{S11}$$

where $k = (\mathbf{k},\omega)$ labels both momentum and energy, $G^0(k)$ denotes the Green's function for the noninteracting system (*S7*), and

$$\Sigma^*(k) = \int dp\, G(p+k)V^*(p)\Gamma(k+p,k,p) \tag{S12}$$

is the full, interacting, quasiparticle self-energy. Eq. S12, which ignores divergent contributions from a uniform background of charge, is exact and is defined in terms of the screened Coulomb interaction,

$$V^*(k) = \frac{V(k)}{1-V(k)\Pi(k)} = \frac{V(k)}{\varepsilon(k)}, \tag{S13}$$

where $V(k) = 2\pi e^2/k$ is the bare Coulomb interaction and $\varepsilon(k)$ is the frequency- and momentum-dependent dielectric function. $\Gamma(k_1,k_2,k_3)$ is the vertex function.

The usual perturbation framework in quantum electrodynamics (QED) would be to expand $G$, $V^*$, and $\Gamma$ in powers of the fine structure constant, which in a Dirac system has the value $\alpha_g = e^2/\hbar v_F$. However, because $\alpha_g = 2.2$ in suspended graphene, such a perturbation series will not converge except perhaps after a summation of an infinite number of terms, to all orders in $\alpha_g$.

Our IXS experiments suggest that the Dirac fermions in graphene are more polarizable than originally expected. Specifically, at small $\omega$ and $\mathbf{k}$, with $\omega < \hbar v_F k$, the dielectric function $\varepsilon \to 15$, which is 3.5 times larger than estimates from RPA (*28*). A better controlled procedure for computing $\Sigma^*(\mathbf{k},\omega)$, as well as other physical observables, would then be to expand in powers of the screened interaction, $V^*$ (Eq. S13). The expansion for the self-energy, for example, may be expressed diagrammatically as

$$\Sigma^* = \text{[diagram]} = \text{[diagram]} + \text{[diagram]} + \text{[diagram]} + o[(\alpha^*)^3]$$

. (S14)

In Eq. S14, a double-wavy line represents the screened Coulomb interaction, $V^*$, a solid line represents the bare Green's function $G^0$, a double solid line is the full Green's function, $G$, and the filled triangle represents the full vertex function, $\Gamma$. In a Dirac system, each screened Coulomb line introduces a factor of

$$\alpha_g^*(\mathbf{k},\omega) = \frac{e^2}{\hbar v_F \varepsilon(\mathbf{k},\omega)},$$
(S15)

which resides within the integrand of each diagram. In other words, the perturbation expansion for $\Sigma^*(\mathbf{k},\omega)$, as well as any other quantity that depends on internal Coulomb lines, may be organized into powers of $\alpha_g^*$, rather than the bare $\alpha_g$.

Because $\alpha_g^*(\mathbf{k},\omega)/\alpha_g \leq 1$ at infrared frequencies for all momenta, and $\alpha_g^*(\mathbf{k},\omega)/\alpha_g \ll 1$ for $\omega < \hbar v_F k$ as $k \to 0$, the series expansion for $\Sigma^*(\mathbf{k},\omega)$ in powers of $\alpha_g^*(\mathbf{k},\omega)$ should be more convergent than the expansion in terms of the bare $\alpha_g$. As a consequence, the interaction between dressed quasiparticles, which is captured by the screened $\alpha_g^*(\mathbf{k},\omega)$, should be on the whole weaker than interactions between the bare electrons in the Dirac portion of the spectrum.

We emphasize that the screened coupling $\alpha_g^*(\mathbf{k},\omega)$ is not a "background" parameter, but the result of the exact resummation of an infinite class of diagrams to all orders in the bare coupling constant $\alpha_g$. Therefore, $\alpha_g^*$ cannot be used in one-to-one correspondence with $\alpha_g$. For example, it would be incorrect to use $\alpha_g^*$ as input into *ab initio* electronic structure calculations, such as density functional theory. This would result in an erroneous double-counting of the Coulomb interaction.

It is worth pointing out that in the standard, relativistic quantum electrodynamics in 3+1 dimensions, the Green's functions and vertex functions are expressed as power series expansions in the renormalized coupling constant, $\alpha$, which is a scale-dependent quantity approaching at low energies and low momenta the experimentally determined value $\alpha = 1/137$ (*S8*). By analogy, in graphene, expansion of physical quantities to lowest order in the bare $\alpha_g$ neglects the physical fact that, in an interacting system, the effective coupling constant is not "protected," and is in fact renormalized by vacuum polarization effects ranging from low through high energy scales.

**Supporting Figures**

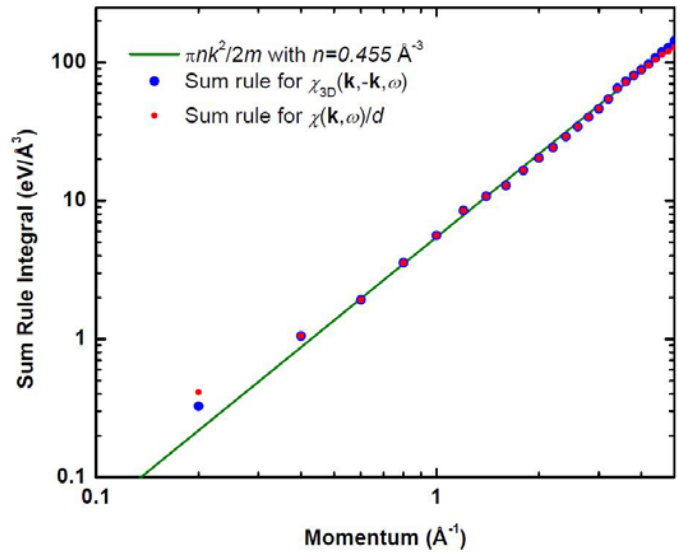

**Figure S1** Value of the sum rule integral in Eq. S1, showing agreement over 1.5 decades of momenta. Note that the sum rule is still satisfied after application of Eq. S9, indicating that the transformation is unitary with respect to spectral weight. At $k > 8$ Å$^{-1}$ deviation from the sum rule is observed because of the emergence of Compton scattering.

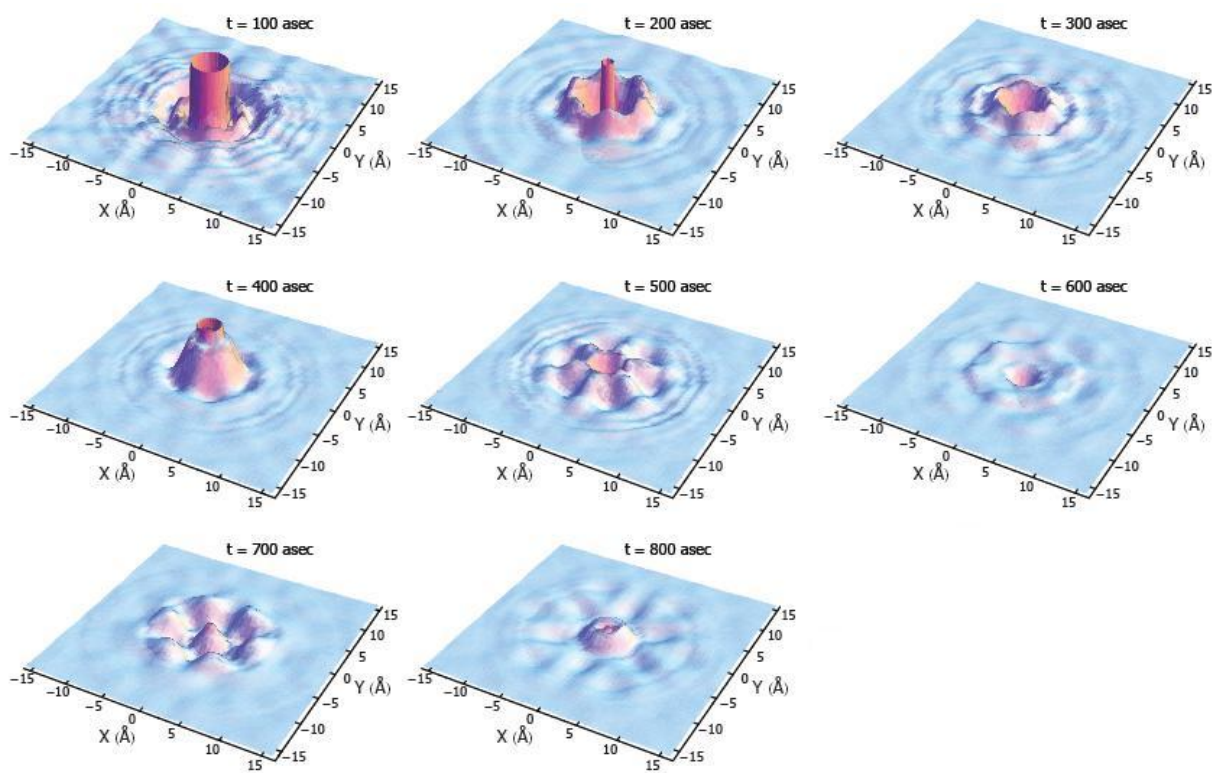

**Figure S2** Dynamical electron density in graphite, $n_{ind}(x, y, t)$, induced by a point source appearing at the origin and instantaneously disappearing, plotted over a field of view of 30 Å × 30 Å in time steps of 100 as ($10^{-16}$ sec). The resolutions for these images were $\Delta r$ = 0.20 Å and $\Delta t$ = 10.3 as. Note that this time resolution is faster than any pulses currently available from ultrafast lasers. These images were generated assuming a homogeneous distribution of charge (see text).

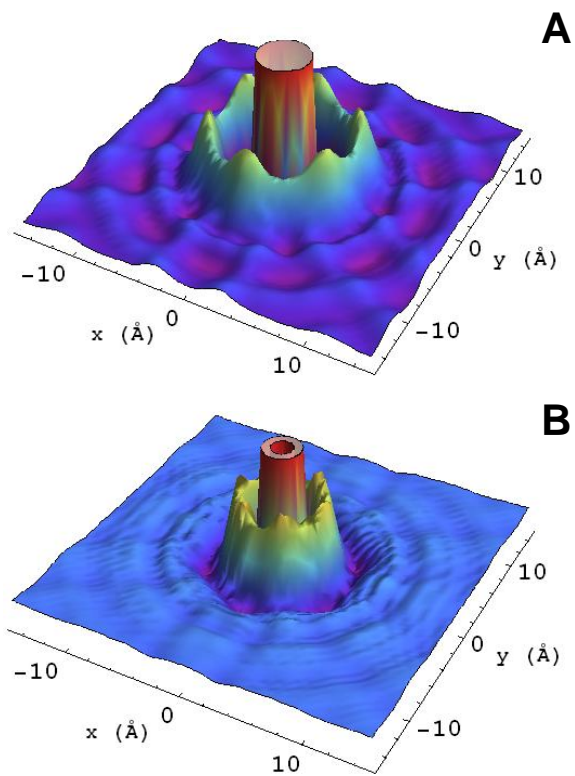

**Figure S3** Images of the electron density, $n_{ind}(x, y)$, induced by a static point charge in graphite, determined (a) in the homogeneous limit and (b) in the case of an arbitrarily thin layer (see text). The six-fold angular modulation arises from the anisotropic dispersion of the $\pi$ plasmon.

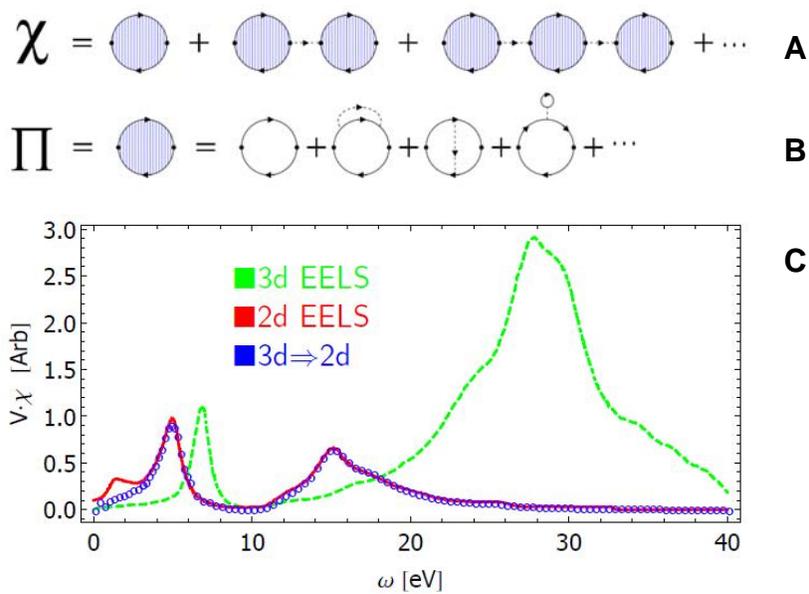

**Figure S4** Schematic representation of (a) the response function and (b) the polarization function. (c) Test of the conversion expression Eq. S9 on the theoretical calculations of Eberlein (*24*). The green curve is their calculation of the response function of graphite, the red curve is their corresponding calculation for graphene. The blue circles are the curve acquired by applying Eq. S9 to their graphite calculation, which agrees quite well with the calculation for graphene (see text).

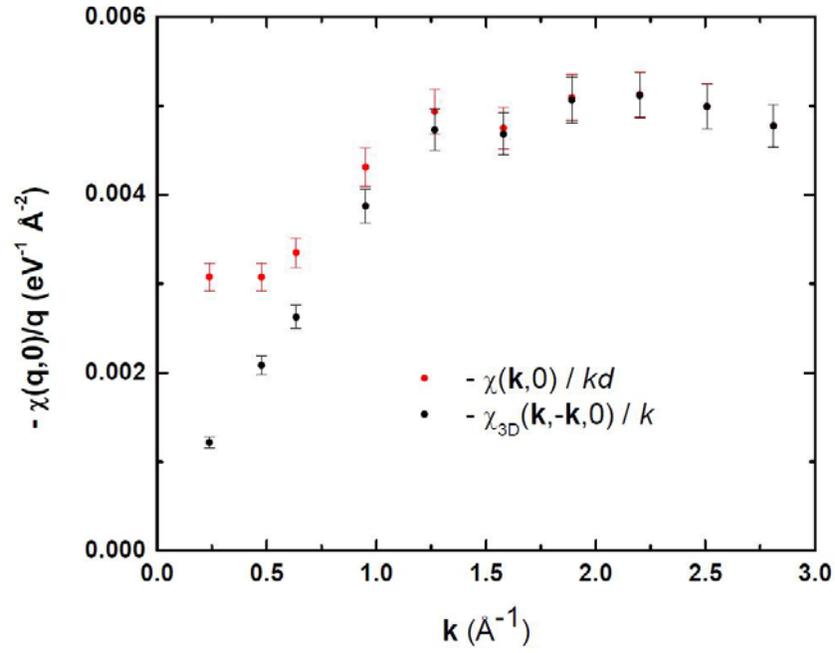

**Figure S5** Long wavelength asymptotic properties of the static response for both graphite and graphene, plotted against momentum. The values are scaled to the momentum value to reveal the leading behavior. Unlike $\chi_{3D}(\mathbf{k},-\mathbf{k},0)$, the graphene response is linear in **k**, creating dielectric screening in two dimensions.

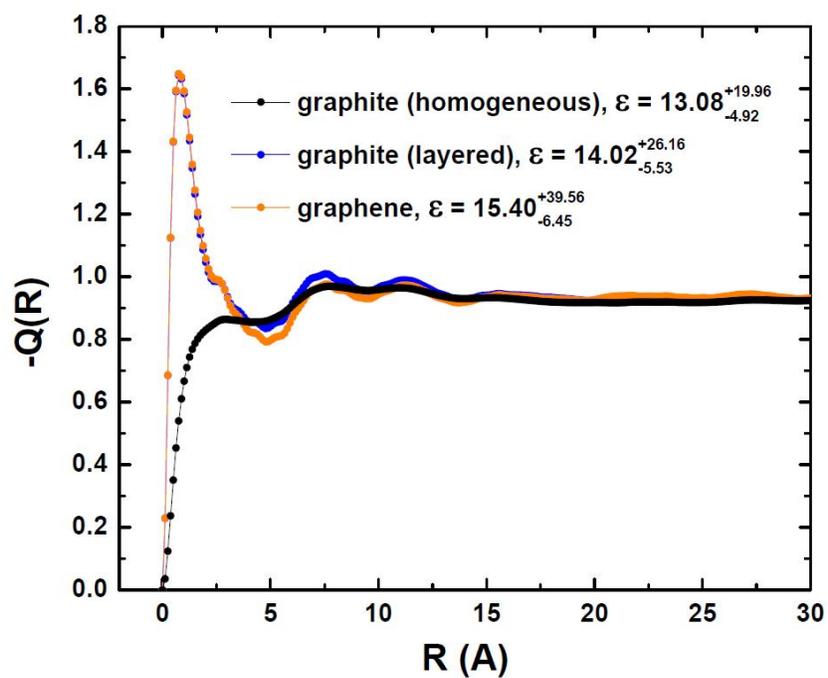

**Figure S6** Accumulated charge, $Q(R)$, plotted against the integration cutoff, $R$, for the for graphite in the homogeneous and layered limits, as well as graphene. The error bars on the dielectric constants come from a assuming a 5% deviation from the F sum rule.